\documentclass[%
 reprint,
 amsmath,amssymb,
 aps,
]{revtex4-2}

\usepackage{graphicx}
\usepackage{dcolumn}
\usepackage{bm}
\usepackage{xcolor}
\usepackage{hyperref}

\begin{document}

\preprint{APS/123-QED}

\title{Negative Compressability of Non-Equilibrium Non-Ideal Bose--Einstein Condensate}

\author{Vladislav Yu. Shishkov}
 \email{vladislavmipt@gmail.com}
 \affiliation{Dukhov Research Institute of Automatics (VNIIA), 22 Sushchevskaya, Moscow 127055, Russia;}
 \affiliation{Moscow Institute of Physics and Technology, 9 Institutskiy pereulok, Dolgoprudny 141700, Moscow region, Russia;}
 \affiliation{ Center for Photonics and Quantum Materials, Skolkovo Institute of Science and Technology, Moscow, Russian Federation; }
 \affiliation{ Laboratories for Hybrid Photonics, Skolkovo Institute of Science and Technology, Moscow, Russian Federation; }
\author{Evgeny S. Andrianov}%
 \affiliation{Dukhov Research Institute of Automatics (VNIIA), 22 Sushchevskaya, Moscow 127055, Russia;}
 \affiliation{Moscow Institute of Physics and Technology, 9 Institutskiy pereulok, Dolgoprudny 141700, Moscow region, Russia;}
 \affiliation{ Center for Photonics and Quantum Materials, Skolkovo Institute of Science and Technology, Moscow, Russian Federation; }
 \affiliation{ Laboratories for Hybrid Photonics, Skolkovo Institute of Science and Technology, Moscow, Russian Federation; }

\date{\today}

\begin{abstract}
An ideal equilibrium Bose--Einstein condensate (BEC) is usually considered in the grand canonical ($\mu V T$) ensemble, which implies the presence of the chemical equilibrium with the environment.
However, in most experimental scenarios, the total amount of particles in BEC is determined either by the initial conditions or by the balance between dissipation and pumping.
As a result, BEC may possess the thermal equilibrium but almost never the chemical equilibrium.
In addition, many experimentally achievable BECs are non-ideal due to interaction between particles.
In the recent work [10.1103/PhysRevLett.128.065301], it has been shown that invariant subspaces in the system Hilbert space appear in non-equilibrium BEC in the fast thermalization limit.
In each of these subspaces, Gibbs distribution is established with a certain number of particles that makes it possible to investigate properties of non-ideal non-equilibrium BEC independently in each invariant subspace.
In this work, we analyze the BEC stability due to change in dispersion curve caused by non-ideal interactions in BEC.
Generally, non-ideal interactions lead to the redshift or blueshift of the dispertion curve and to the change in the effective mass of the particles.
We show that the redshift of the dispersion curve can lead to the negative compressibility of BEC, whereas the change in the effective mass always makes BEC more stable.
We find the explicit condition for the particle density in BEC, at which the negative compressibility appears.
\end{abstract}

\maketitle

\section{Introduction}
The first Bose--Einstein condensates (BECs) were experimentally demonstrated in 1995~\cite{anderson1995observation, davis1995bose}, almost 70 years after the theoretical prediction~\cite{bose1924plancks, einstein1925quantentheorie}. 
Nowadays, BECs have a wide range of applications in low-energy optoelectronics~\cite{sanvitto2016road, keeling2020bose, kavokin2022polariton}, including fast optical switching operating at quantum limit~\cite{zasedatelev2021single}, as well as driving chemical reactions~\cite{pannir2022driving}.
Besides practical applications, recently, a great interest has been attracted to the fundamental problems concerning physical properties of BECs.
One of such fundamental problem is the stability of BEC.
Several different mechanisms for instability formation in BEC such as mechanical collapse~\cite{sackett1999measurements, timmermans1999rarified, nath2009phonon, bortolotti2006scattering, buljan2005incoherent, donley2001dynamics, debnath2010instability, jeon2002pairing, miyakawa2001induced, mueller2000finite, eleftheriou2000instability, stoof1994atomic}, dynamical instability~\cite{van2007dynamical, saito2002split, berman2002quantum, hanai2017dynamical, bobrovska2018dynamical, baboux2018unstable, baharian2013bose, kanamoto2003quantum, wamba2013instability}, and modulational instability~\cite{xue2005modulational, baizakov2002regular, kourakis2005modulational, robins2001modulational, konotop2002modulational, theocharis2003modulational, jin2005modulational, doktorov2007full, everitt2017observation, ferrier2018onset, molignini2018superlattice, qi2012modulational, wamba2014dynamical, tamilthiruvalluvar2019impact, bhat2021modulational, wamba2008variational} have been proposed and experimentally verified.

In many experimental implementations, BEC may be in temperature equilibrium, but the chemical equilibrium is almost never established.
For example, in BECs formed by trapped atoms, the total number of particles is determined by the initial conditions and does not change as the system comes to thermal equilibrium~\cite{griffin1996bose, andrews1997observation, davis1995bose}.
In BECs based on excitons and polaritons, there is no even thermal equilibrium because the total number of particles is determined by the balance of the rates of external pumping and losses, as, for example, in polariton BECs~\cite{deng2003polariton, kasprzak2006bose, combescot2015excitons, byrnes2014exciton, wertz2010spontaneous, balili2007bose, estrecho2018single, klaas2018photon, sun2017bose, deng2010exciton, klaas2018photon, imamog1996nonequilibrium, keeling2020bose, wei2019low, zasedatelev2019room, plumhof2014room}.

In most realizations, BEC is not only non-equilibrium but also is non-ideal due to the presence of non-linear effects.
The nature of non-ideality can be different depending on the implementation of the condensate.
In inorganic semiconductors, the non-ideality of polariton BECs is caused by the Coulomb interaction between exciton components.
This interaction leads to the depletion of the condensate and the appearance of a linear region in its dispersion curve.
For polariton BECs based on organic dyes~\cite{deng2003polariton, kasprzak2006bose, combescot2015excitons, byrnes2014exciton, wertz2010spontaneous, balili2007bose, estrecho2018single, klaas2018photon, sun2017bose, deng2010exciton, klaas2018photon, imamog1996nonequilibrium, keeling2020bose, wei2019low, zasedatelev2019room, plumhof2014room}, non-ideality has a different nature since the spatial localization of Frenkel excitons suppresses the Coulomb interaction.
The non-linearity of such polaritons is associated with the saturation of the exciton component, which leads to the change in the permittivity~\cite{yagafarov2020mechanisms, schwab2021mechanisms}. 
In this case, non-linearity does not lead to a significant distortion of the dispersion curve of polaritons, i.e. no linear region in the vicinity of ${\bf k}={\bf 0}$ appears~\cite{yagafarov2020mechanisms, sabatini2020organic, putintsev2020nano, dusel2021room, cookson2017yellow, sannikov2019room, wei2021low, plumhof2014room, daskalakis2014nonlinear}. 
Instead, the dispersion curve shifts as a whole.

In this regard, the important problem of the influence of non-ideal and non-equilibrium nature of BEC on such macroscopic characteristics as stability and compressibility arises. 
The problem has so far remained difficult for investigation because the consistent description of BEC in the grand canonical ensemble ($\mu VT$-ensemble) is impossible, since the latter assumes both temperature and chemical equilibrium between the condensate and the environment which may be absent in many situations.
For these reasons, the BEC description requires the first principles methods such as the Gross--Pitaevskii equation~\cite{carusotto2013quantum}, the Lindblad equation~\cite{kavokin2017microcavities, sanvitto2012exciton, laussy2004spontaneousPSSC, laussy2004spontaneousPRL}, and the Maxwell--Boltzmann equations~\cite{malpuech2002room, banyai2002real, cao2004condensation, doan2008coherence, tassone1997bottleneck, kirton2013nonequilibrium, kirton2015thermalization, strashko2018organic, arnardottir2020multimode}.
However, it is important to have the analytical expressions of the influence of non-ideality on the properties of a non-equilibrium BEC.

Generally, the thermalization process by itself leaves the total number of particles unchanged in non-equilibrium BECs~\cite{sanvitto2012exciton}.
Thus, the total number of particles is the integral of motion for this relaxation process.
Due to the presence of the integral of motion, the invariant subspaces appear in the system~\cite{shishkov2018zeroth}.
These invariant subspaces are characterized by a certain number of particles distributed in the system as a whole~\cite{shishkov2018zeroth, shishkov2021exact}.
In the fast thermalization limit, a Gibbs distribution with an equilibrium temperature is established in each invariant subspace regardless of other relaxation processes~\cite{shishkov2021exact, shishkov2021fully}.
The Gibbs distribution with the fixed total number of particles corresponds to the canonical ensemble ($NVT$ ensemble).
Thus, in the case of fast thermalization, the $NVT$ ensemble plays a key role in the dynamics of non-equilibrium BEC and can make it possible to analyze BEC stability and obtain analytical estimates for compressibility.

In this paper, we consider non-ideal two-dimensional BEC in a canonical ($NVT$) ensemble.
We find the compressibility of BEC and show that the compressibility of an ideal BEC always remains positive and increases with the number of particles in the condensate.
For a non-ideal BEC, when the non-linearity leads to a redshift, a negative compressibility of the condensate may occur.
Negative compressibility indicates the instability of a non-ideal BEC.
We also show that the change in the effective mass always makes BEC more stable.

\section{BEC non-ideality and change in the dispersion curve}
We consider a non-ideal 2D Bose gas localized in the region $V$ ($V$ has the dimension of area) with the degeneracy $g$ of each state.
Such a non-ideal Bose gas is described by the Hamiltonian
\begin{equation} \label{H}
\hat H = \hat H^{\rm ideal} + \hat H^\text{non-ideal}
\end{equation}
where $\hat H^{\rm ideal}$ is the ideal part and $\hat H^\text{non-ideal}$ is the non-ideal part of the Hamiltonian.

The ideal part of the Hamiltonian, $\hat H^{\rm ideal}$, has the form
\begin{equation} \label{H_ideal}
\hat H^{\rm ideal} = \sum_{{\bf k},\lambda} \hbar \omega_{\bf k} \hat a^\dag_{{\bf k},\lambda} \hat a_{{\bf k},\lambda},
\end{equation}
where ${\bf k}$ is the wave vector of Bose gas particles, the frequency $\omega_{\bf k}$ corresponds to this wave vector, which is calculated according to $\omega_{\bf k} = \omega_{\bf 0} + \alpha {\bf k}^2$, and $\lambda$ enumerates different states with the same wave vector.

\begin{figure}
    \includegraphics[width=0.85\linewidth]{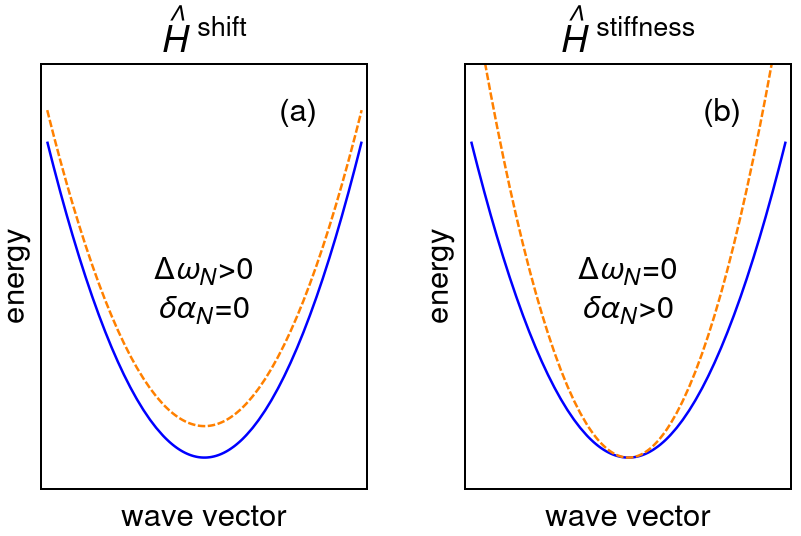}
    \caption{
Change in the dispersion curve due to non-linearity.
The blue solid line denotes the original dispersion curve, the orange dotted line shows the change in the dispersion curve due to (a) $\hat H^{\rm shift}$ and (b) $\hat H^{\rm stiffness}$.
The change in the dispersion curve corresponds to (a) $\Delta\omega_N>0$, $\delta \alpha_N=0$ ($\varkappa_1>0$, $\varkappa_2=0$) and (b) $\Delta\omega_N= 0$, $\delta \alpha_N>0$ ($\varkappa_1=0$, $\varkappa_2>0$).
}
    \label{fig:dispersion}
\end{figure}

Due to the non-linearity, an increase in the number of particles leads to a change in the dispersion curve.
As noted in the Introduction, in polariton systems based on organic dyes, the non-linearity is associated with a change in the permittivity~\cite{yagafarov2020mechanisms, schwab2021mechanisms}.
In this case, changes in the dispersion curve near ${\bf k}={\bf 0}$ can be divided into two parts: change in the energy of the ground state of polaritons (the energy of each state of polaritons changes by the same value) and change in the effective mass of polaritons (Fig.~\ref{fig:dispersion}).
The non-linear part of the Hamiltonian~(\ref{H}) that leads to such changes in the dispersion curve is
\begin{equation} \label{H_non-ideal}
\hat H^\text{non-ideal}  = \hat H^{\rm shift} + \hat H^{\rm stiffness}
\end{equation}
with
\begin{equation} \label{H_shift}
\hat H^{\rm shift} = \frac{\hbar \varkappa_1}{2V} \hat N (\hat N - 1)
\end{equation}
\begin{equation} \label{H_stiffness}
\hat H^{\rm stiffness} = \frac{\hbar \varkappa_2}{V} (\hat N - 1) \sum_{{\bf k},\lambda} \alpha {\bf k}^2 \hat a_{{\bf k},\lambda}^\dag \hat a_{{\bf k},\lambda}
\end{equation}
where $\hat N$ is the operator of the total number of particles distributed in the system as a whole, $\varkappa_1$ and $\varkappa_2$ characterize the non-ideality of the Bose gas. 
We assume that $\varkappa_1$ and $\varkappa_2$ do not depend on $V$.

The Hamiltonians $\hat H^{\rm shift}$ and $\hat H^{\rm stiffness}$ lead to the shift in the energy of the ground state of the system and to the change in the effective mass of particles, as shown in Fig.~\ref{fig:dispersion}.
Indeed, the Heisenberg equation for $\hat a_{{\bf k},\lambda}$ has the form
\begin{multline} \label{a_heizenberg}
\hat a_{{\bf k},\lambda} = 
-i 
\left[
\left( \omega_{\bf k} + \frac{\varkappa_1 \hat N}{V} \right) + \left( 1 + \frac{\varkappa_2 \hat N}{V} \right) \alpha {\bf k}^2
\right]
\hat a_{{\bf k},\lambda} \\
-i \frac{\varkappa_2}{V} \sum_{{\bf q},\lambda'} \alpha {\bf q}^2 \hat a_{{\bf q},\lambda'}^\dag \hat a_{{\bf q},\lambda'} \hat a_{{\bf k},\lambda}
\end{multline}
For a fixed number of particles $N$, the non-linearity $\hat H^{\rm shift}$ leads to a shift of the entire dispersion curve by the frequency
\begin{equation} \label{frequency_shift}
\Delta \omega_N = \frac{\varkappa_1 N}{V}.
\end{equation}
This shift is proportional to the particle density, which is typical for BEC implementations based on polaritons~\cite{yagafarov2020mechanisms, sabatini2020organic, putintsev2020nano, dusel2021room, cookson2017yellow, sannikov2019room, wei2021low, plumhof2014room, daskalakis2014nonlinear}.
For sufficiently large $\bf k$ in BEC, the term $\sum_{{\bf q},\lambda'} {\bf q}^2 \langle \hat a_{{\bf q},\lambda'}^\dag \hat a_{{\bf q},\lambda'} \rangle$ is much less compared to $\langle \hat N \rangle {\bf k}^2$.
In this case, the non-linearity of $\hat H^{\rm stiffness}$ leads to a relative change in the effective mass
\begin{equation} \label{stiffness_change}
\delta \alpha_N = \frac{\varkappa_2 N}{V}.
\end{equation}

Below we analyze the stability of BEC and the Bose gas before condensation in the $NVT$ ensemble.
$NVT$ ensemble corresponds to the experimental realization of BEC, where the condensate does not exchange the particles with the environment.
As it was discussed in the Introduction, this situation is mostly realized in experiments with trapped atoms.
However, $NVT$ ensemble also plays an important role in the evolution of the non-equilibrium BEC, when the condensate dissipates, but remains sustained due to an external pumping.
The appropriate way to describe such condensates is to use the Lindblad master equation~\cite{kavokin2017microcavities}.
This equation reliably takes into account the open nature of the condensates, including the thermalization processes. 
Generally, this equation is difficult to solve due to the great amount of degrees of freedom of the condensate.
One way to overcome this difficulty is to use one of the mean-field approaches, i.e. Maxwell-Boltzmann equations~\cite{malpuech2002room, banyai2002real, cao2004condensation, doan2008coherence, tassone1997bottleneck, kirton2013nonequilibrium, kirton2015thermalization, strashko2018organic, arnardottir2020multimode}.
Another approach to solve the Lindblad equation has been developed in~\cite{shishkov2021exact, shishkov2021fully}.
There, it has been shown that in the fast thermalization limit the complexity of the Lindblad equations can be substantially reduced due to the presence of the integral of motion in the thermalization process.
Moreover, in this case, the stationary state of non-equilibrium BEC is determined by the corresponding thermodynamical $NVT$ ensemble~\cite{shishkov2021exact, shishkov2021fully}.
Thus, the analysis of BEC stability in an $NVT$ ensemble is relevant not only for equilibrium BEC, but also for non-equilibrium BEC with the fast thermalization.

\section{Partition function of non-ideal BEC}
For definiteness, below we consider the degeneracy ${g=1}$ and ${g=2}$.
The latter situation is typical for BECs based on polaritons, which have two polarizations and, accordingly, each state with a certain wave vector $\bf k$ has degeneracy 2.
The density of states, $\nu$, is
\begin{equation} \label{density of states}
\nu = \frac{gV}{4\pi \hbar \alpha}, \;\; \left[ \nu \right] = {\rm eV}^{-1}
\end{equation}
We assume that ${\nu k_BT \gg 1}$, that is the number of states in the energy interval ${(\hbar\omega_{{\bf k}={\bf 0}}, \hbar\omega_{{\bf k}={\bf 0}}+k_BT)}$ is much greater than one.

The partition function, $Z_N^{\rm ideal}$, of an ideal Bose gas in the $NVT$ ensemble is
\begin{equation} \label{Z_ideal}
Z_N^{\rm ideal}(T)={\sum}' 
e^{- \sum_{{\bf k},\lambda} n_{{\bf k},\lambda}\hbar(\omega_{\bf k}-\omega_{\bf 0})/{k_BT}}
\end{equation}
where $T$ is the temperature of the Bose gas, $k_B$ is the Boltzmann constant, $n_{{\bf k}, \lambda}$ is the number of particles in the state with wave vector $\bf k$ and polarization $\lambda$, and ${\sum}'$ is the sum over all possible configurations of particles, provided that total number of particles distributed in the system as a whole equals $N$.
The partition function obeys the recursive relation~\cite{shishkov2021fully}
\begin{equation} \label{Z_reccurence}
Z_N^{\rm ideal}(T) = \sum_{n=0}^{N-1} Z_n^{\rm ideal}(T) \left( \frac{g}{N} + \frac{\nu k_BT}{N(N-n)} \right) 
\end{equation}
where $Z_0^{\rm ideal}(T)=1$.

Depending on the relation between $\nu k_BT$ and $N$, two regions can be separated in the $\{N,V,T\}$ plane: Bose gas is far from the condensation state~($N \ll \nu k_BT$ ), the Bose gas is in the BEC~($N \gg \nu k_BT$) state~\cite{shishkov2021fully}.
From Eq.~(\ref{Z_reccurence}) it follows that
\begin{equation} \label{Z_before}
Z_N^{\rm ideal}(T) \approx \frac{\left( \nu k_BT \right)^N}{N!} 
\end{equation}
before BEC formation~($N \ll \nu k_BT$) and
\begin{equation} \label{Z_after_g=1}
Z_N^{\rm ideal}(T) \approx \left( 1 - \frac{\nu k_BT}{N}  \right) e^{\pi^2 \nu k_BT/6 }, \; {\rm for} \, g=1
\end{equation}
\begin{equation} \label{Z_after_g=2}
Z_N^{\rm ideal}(T) \approx \left( N - \nu k_BT\ln N \right)  e^{\pi^2 \nu k_BT/6}, \; {\rm for} \, g=2
\end{equation}
after BEC formation~($N \gg \nu k_BT$)~\cite{shishkov2021fully}.

The partition function in the $NVT$ ensemble of the non-ideal BEC with the non-linearity~(\ref{H_non-ideal}) is
\begin{multline} \label{Z}
Z_N(T)={\sum}' 
e^{- \sum_{{\bf k},\lambda} n_{{\bf k},\lambda}\hbar(\omega_{\bf k}-\omega_{\bf 0})/{k_BT}}
\times \\ 
e^{\hbar\Delta\omega_N(N-1)/{2k_BT}}
e^{- \delta\alpha_N (N-1) \sum_{{\bf k},\lambda} n_{{\bf k},\lambda}\hbar(\omega_{\bf k}-\omega_{\bf 0})/{Nk_BT}}
\end{multline}
Comparing the expressions~(\ref{Z_ideal})~and~(\ref{Z}), it is easy to show that 
\begin{equation} \label{Z_N}
Z_N(T)  
= Z_N^{\rm ideal}\left( \frac{T}{1+(1-N^{-1}) \delta \alpha_N} \right) e^{-\hbar\Delta\omega_N(N-1)/2k_B T}
\end{equation}

The explicit expressions for the partition functions~(\ref{Z_before})--(\ref{Z_after_g=2}) and (\ref{Z_N}) enable to find the compressibility of the non-ideal BEC and investigate its stability.

\section{COMPRESSIBILITY OF A NON-IDEAL BEC}
The compressibility, $\beta_T$, can be found according to $\beta_T = -\left( Vk_BT {\partial^2 \ln Z_N / \partial V^2} \right)^{-1}$~\cite{landau2013statistical}.
The compressibility of the non-ideal Bose gas before the condensate formation~($N \ll \nu k_BT / (1 + \delta \alpha_N)$) is (see Eq.~(\ref{Z_before})) 
\begin{equation} \label{compressibility_before}
\beta_T \approx \frac{V}{N k_BT } \left(2 - \frac{1}{(1+\delta \alpha_N)^2} + \frac{\hbar\Delta\omega_N}{k_BT} \right)^{-1}.
\end{equation}
After the formation of BEC~($N \gg \nu k_BT / (1 + \delta \alpha_N)$), the compressibility becomes (see Eqs.~(\ref{Z_after_g=1})--(\ref{Z_after_g=2})) 
\begin{multline} \label{compressibility_after_g=1}
\beta_T \approx \frac{V}{k_BT} 
\left[
\left( \frac{\nu k_B T}{N} \right)^2 \frac{(1+2\delta \alpha_N)^2}{(1+\delta \alpha_N)^4} + 
\frac{\hbar\Delta\omega_N}{k_B T}N +
\right.  
\\
\left.
+ \left(2 + \frac{\pi^2}{6}\right) \frac{\nu k_B T}{N} \frac{\delta \alpha_N^2}{(1+\delta \alpha_N)^3}  
\right]^{-1}
, \; {\rm for} \, g=1,
\end{multline}
\begin{multline} \label{compressibility_after_g=2}
\beta_T \approx \frac{V}{k_BT} 
\left[
\left( \frac{\nu k_B T \ln N}{N} \right)^2 \frac{(1+2\delta \alpha_N)^2}{(1+\delta \alpha_N)^4} + 
\frac{\hbar\Delta\omega_N}{k_B T}N +
\right.  
\\
\left.
+ \left(2 + \frac{\pi^2}{6}\right) \frac{\nu k_B T \ln N}{N} \frac{\delta \alpha_N^2}{(1+\delta \alpha_N)^3}  
\right]^{-1}
, \; {\rm for} \, g=2,
\end{multline}

For the ideal BEC, when $\Delta\omega_N=0$ and $\delta \alpha_N = 0$, the compressibility is always positive, and, hence, the ideal BEC is stable.
However, when ideal BEC is formed, $\beta_T\propto N^2$. 
Therefore, as the the total number of particles grows, the volume occupied by the BEC becomes more sensitive to the changes in pressure.
This situation is exactly opposed to the behaviour of an ideal Bose gas before the BEC formation.
In the latter case, $\beta_T\propto N^{-1}$ and an increase in the number of particles leads to a decrease in the compressibility.
Thus, above the condensation threshold, as the total number of particles increases the volume fluctuations strongly increase and BEC becomes less stable.
Indeed, the square of the thermal fluctuations of the volume, $(\Delta V)^2$, is proportional to the compressibility of $\beta_T$, so ${(\Delta V)^2 = Vk_BT\beta_T}$~\cite{landau2013statistical}.

\begin{figure}
    \includegraphics[width=\linewidth]{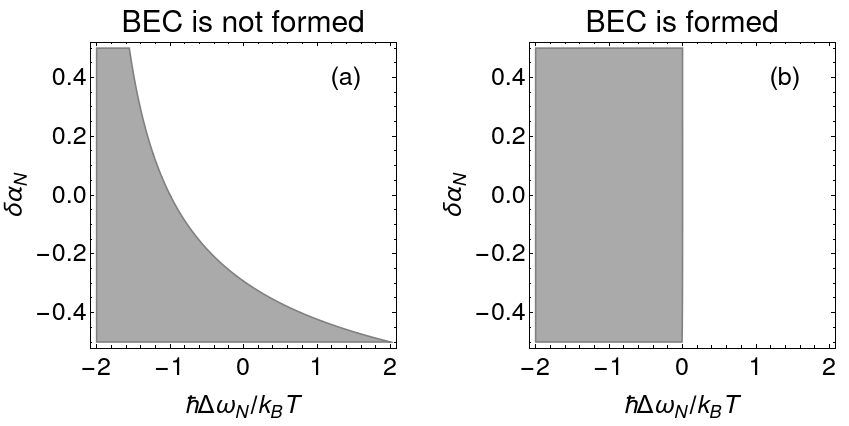}
    \caption{
Polariton stability diagram below (a) and above (b) the BEC formation (${g=1}$).
Unstable regions are shaded with gray and correspond to the negative isothermal compressibility.
Figure (b) corresponds to~${N=10^3\nu k_BT}$.
}
    \label{fig:stability_regions}
\end{figure}

Let us consider the case when the non-linearity does not lead to a change of the effective mass~($\delta\alpha_N=0$).
In this case, the compressibility is always positive at blueshift ($\Delta\omega_N > 0$).
However, in the case of redshift ($\Delta\omega_N < 0$), non-ideality can lead to the negative compressibility and instability~(Fig.~\ref{fig:stability_regions}).
In this case, there is a certain critical number of particles, at which the instability emerges.
We denote this critical number of particles for a Bose gas below the condensation threshold as $N_c^{\rm Bose}$ and above the condensation threshold as $N_c^{\rm BEC}$.
Form Eqs.~(\ref{compressibility_before})--(\ref{compressibility_after_g=2}) it follows that
\begin{equation} \label{nc_bose}
N_c^{\rm Bose} = \frac{V k_BT}{\hbar| \varkappa_1 |} \; \Longleftrightarrow \; \hbar\Delta \omega_{N_c^{\rm Bose}} = -k_BT
\end{equation}
\begin{equation} \label{nc_bec_g=1}
N_c^{\rm BEC} = \left( N_c^{\rm Bose} \right)^{1/4} \left(\nu k_BT\right)^{1/2}, \; {\rm for} \, g=1
\end{equation}
\begin{equation} \label{nc_bec_g=2}
\frac{N_c^{\rm BEC}}{\sqrt{ \ln N_c^{\rm BEC} }} = \left( N_c^{\rm Bose} \right)^{1/4} \left(\nu k_BT\right)^{1/2} , \; {\rm for} \, g=2
\end{equation}
Thus, instability can appear both in non-ideal BEC and non-condensed Bose gas.
In a non-condensed Bose gas, instability occurs when the frequency redshift reaches $k_BT/\hbar$ (see equations~(\ref{nc_bose})~and~(\ref{frequency_shift})).
In BEC, instability arises at a much smaller number of particles.
Indeed, from the expressions~(\ref{nc_bec_g=1})--(\ref{nc_bec_g=2}) it follows that $N_c^{\rm BEC} \ll N_c^{\rm Bose}$ for $\nu k_BT \ll N_c^{\rm Bose}$.
Moreover, ${N_c^{\rm BEC} \propto V^{3/4}}$, therefore, the critical concentration of particles, $N/V$, at which instability occurs in BEC falls as $V^{-1/4}$, whereas, in a non-condensed Bose gas, the critical concentration does not depend on~$V$.

A change in the effective mass of the particles has a different effect on the stability of the system before and after BEC formation.
Before the BEC formation, the change in the dispersion curve strongly affects the stability of the system: if the non-ideality leads to an increase in the effective mass ($\delta \alpha_N>0$), then a larger redshift $\Delta \omega_N$ is required to reach the instability, but, in the opposite case ($\delta \alpha_N<0$), the stability region with respect to $\Delta \omega_N$ narrows~(Fig.~\ref{fig:stability_regions}(a)).
This behavior is expected.
Indeed, with the increase in the effective mass, the additional pressure due to non-ideality is positive, thus, the system becomes more stable.
However, if the effective mass decreases with the increase of the number of particles, then the corresponding additional pressure is negative, thus, the system becomes more unstable.

The change in the effective mass has a less significant effect on the BEC stability~(Fig.~\ref{fig:stability_regions}(b)).
This is because the change in the shape of the dispersion curve leads to two opposite effects, that almost compensate each other.
For instanse, let us consider $\delta \alpha_N>0$.
On the one hand, at $\delta \alpha_N>0$, the effective mass increases, which leads to the increase in pressure.
On the other hand, $\nu$ decreases, therefore, the fraction of particles in the ground state rises, that reduces the pressure.
A more detailed analysis of Eqs.~(\ref{compressibility_after_g=1})--(\ref{compressibility_after_g=2}) shows that, regardless of the sign of $\delta \alpha_N$, the final additional pressure is always positive in BEC.

\section{Conclusion}
We considered a non-ideal two-dimensional Bose--Einstein condensate (BEC) in a canonical ensemble, when the non-ideality leads to a change in the dispersion curve, namely, an increase in the effective mass and a shift of the dispersion curve.
We obtained the explicit expression for the partition function of the two-dimensional non-ideal BEC, which made it possible to study compressibility.
We demonstrated that a change in the effective mass due to the non-ideality has little effect on stability of BEC.
We also showed that when non-linear interactions in BEC result in a blueshift of the dispersion curve, the compressibility always remains positive.
However, in the case of redshift, the BEC compressibility becomes negative at a sufficiently high particle concentration.
Thus, the BEC can become unstable at redshift because negative compressibility leads to pressure fluctuations arising in different parts of the BEC are not compensated by the surrounding condensate, but, on the contrary, begin to grow with time.
One can suppose that this may lead to a non-uniform distribution of BEC particles.

The instability of Bose gas at redshift occurs both in the BEC and in the Bose gas prior to condensation.
In the latter case, the instability occurs when the frequency redshift exceeds $k_BT/\hbar$.
However, the BEC formation significantly reduces the concentration of particles required for the onset of the instability.
Moreover, the critical concentration of the particles before the formation of a BEC does not depend on the occupied region, but after the formation of a BEC, the critical concentration of particles decreases with an increase in the size of the condensate.

\begin{acknowledgments}
The research was financially supported by a grant from Russian Science Foundation (project No. 20-72-10145).
E.S.A. and V.Yu.Sh. thank the Foundation for the Advancement of Theoretical Physics and Mathematics ``Basis''.
\end{acknowledgments}

\bibliography{NVT-BEC}

\end{document}